\documentclass[twoside,twocolumn,fleqn]{article}
\usepackage{epsf}
\usepackage{latexsym}
\usepackage{espcrc2}

\newcommand{\beq}[1]{\begin{equation}\label{#1}}
\newcommand{\eeq}{\end{equation}}
\newcommand{\bea}[1]{\begin{eqnarray}\label{#1}}
\newcommand{\eea}{\end{eqnarray}}
\newcommand{\rf}[1]{(\ref{#1})}
\newcommand{\vev}[1]{ {\langle #1 \rangle} }

\begin{document}
{\onecolumn

\title{The Quantum Spacetime of $c>0$ $2d$ Gravity}

\author{J.~Ambj{\o}rn, {\bf K.~N.~Anagnostopoulos} 
\address{The Niels Bohr Institute,
Blegdamsvej 17, DK-2100 Copenhagen \O, Denmark}
and G.~Thorleifsson \address{Facult\"{a}t f\"{u}r Physik
Universit\"{a}t Bielefeld, D-33615, Bielefeld, Germany}\\
\vspace{1mm}}

\begin{abstract}
We review recent developments in the understanding of the fractal
properties of quantum spacetime of 2d gravity coupled to $c>0$
conformal matter. In particular we discuss bounds put by numerical
simulations using dynamical triangulations on the value of the
Hausdorff dimension $d_H$ obtained from scaling properties of two point
functions defined in terms of geodesic distance. Further insight to
the fractal structure of spacetime is obtained from the study of the
loop length distribution function which reveals that the $0<c\le 1$
system has similar geometric properties with pure gravity, whereas the
branched polymer structure becomes clear for $c\ge 5$.

\end{abstract}

\maketitle}
\nopagebreak

\section{Introduction}

Two--dimensional quantum gravity has been a very useful laboratory for
the study of interaction between matter and geometry. The structure of
space--time in the presence of matter is the least understood, albeit
one of the most interesting, aspect of the theory.  The introduction
of the transfer matrix formalism \cite{transfer} allows us to study
the case of pure gravity in a satisfactory way. It tells us that the
space--time has a self similar structure at all scales and that its
dimension is {\it dynamical}. Although the underlying manifold is
two--dimensional, the fractal dimension turns out to be four. In the
case where we couple matter to gravity there is a set of predictions
from numerical investigations \cite{syracuse} and theoretical
approaches \cite{deformation,ksw} which do not seem to agree. Using
string field theory or the transfer matrix approach with {\it a
modified definition of geodesic distance} one obtains that the fractal
dimension of space--time is \cite{deformation}
\beq{*1}
d_h=\frac{2}{|\gamma|}=\frac{24}{1-c+\sqrt{(25-c)(1-c)}}\, ,
\eeq
where $\gamma$ is the string susceptibility and $c$ the central charge
of the conformal theory in flat space describing the matter we couple
to gravity. An alternative result can be obtained using the diffusion
of a fermion in the context of Liouville theory \cite{ksw}:
\beq{*2}
d_h=-2\frac{\alpha_1}{\alpha_{-1}}=
2\times\frac{\sqrt{25-c}+\sqrt{49-c}}{\sqrt{25-c}+\sqrt{1-c}}\, .
\eeq
In eq.~\rf{*2}, $\alpha_n$ denotes the gravitational dressing of a
$(n+1,n+1)$ primary spinless conformal field. In the case of $c=-2$
matter, numerical simulations \cite{us} are in favour of the
prediction \rf{*2} and exclude with confidence \rf{*1} (see
Table~\ref{t:1}). For $0<c\le 1$ matter the situation is still not
very clear. We performed numerical simulations of the
$c=1/2,4/5,1,5,8$ systems and we observe that the fractal dimension
computed is very close to four for $c\le 1$ \cite{syracuse}, as is the
case for pure gravity, whereas it becomes close to two when $c\ge 5$,
as one expects from branched polymers. Although the simulations
clearly favour $d_h\approx 4$ for $0<c\le 1$, the predictions form
eq.~\rf{*2} are not too far from the results to be completely
disproved.  Our simulations clearly show that the self similar
structure found in the case of pure gravity is identical for the
$0<c\le 1$ systems but disappears for $c\ge 5$.
\begin{table*}[hbt]
\setlength{\tabcolsep}{1.5pc}
\newlength{\digitwidth} \settowidth{\digitwidth}{\rm 0}
\catcode`?=\active \def?{\kern\digitwidth}
\caption{The fractal dimension of all $c\le1$ models studied.}
\label{t:1}
\begin{tabular*}{\textwidth}{@{}l@{\extracolsep{\fill}}rrrrrr}
\hline
\multicolumn{6}{c}{$d_h$}\\
\hline
\multicolumn{1}{l}{$c=-2$}        &
\multicolumn{1}{r}{$c=0$}        & 
\multicolumn{1}{r}{$c=1/2$}        & 
\multicolumn{1}{r}{$c=4/5$}        &
\multicolumn{1}{r}{$c=1$}        & 
\multicolumn{1}{c}{Method} \\
\hline
2         &  4           &  6           &  10          & $\infty$ &
theory,  Eq.~\protect\rf{*1} \\
3.562     &  4           &  4.21        &  4.42        & 4.83     & theory,  Eq.~\protect\rf{*2} \\
3.574(8)  & 4.05(15)     & 4.11(10)     & 4.01(9)      & 3.8--4.0 & $n_1(r;N)$  ~~~~FSS\\
3.53--3.60& 3.92{--}4.01 & 3.99{--}4.08 & 3.99{--}4.10 & 3.9--4.1 & $n_1(r;N)$  ~~~~SDS\\
3.59--3.66& 3.85{--}3.98 & 3.96{--}4.14 & 4.05{--}4.20 & 4.0--4.4 & $\vev{l^n(r)}_N$ ~~~~FSS\\
          &              & 4.28(17)     & 4.46(33)     &          & $n_\varphi(r;N)$  ~~~~FSS\\
          &              & 3.90{--}4.16 & 4.00{--}4.30 &          & $n_\varphi(r;N)$  ~~~~SDS\\
          &              & 3.96{--}4.38 & 3.97{--}4.39 &          & $G_\varphi(r)_N$ ~~~~SDS\\
\hline
\end{tabular*}
\end{table*}

\section{RESULTS}

We performed numerical simulations of $c=1/2,4/5,1,5,8$ matter coupled
to $2d$ quantum gravity by means of dynamical triangulations. The
$c=1/2,4/5$ systems were simulated by putting two and three states
Potts model spins on the vertices of the lattice, whereas the
$c=1,5,8$ systems by putting $1,5,8$ Gaussian fields on the vertices
of the lattice. The fractal dimension was defined in terms of
correlators which are functions of the {\it geodesic distance} on the
surface. Two point functions are given by
\bea{*3}
&n_\phi(R,V) = \frac{1}{V} \int {\cal D}g\,{\cal D}\phi \,\, 
\delta(\int\sqrt{g}-V)\, {\rm e}^{-S_M}\nonumber\\
&\,\times\int d^2\xi d^2\xi'\sqrt{g g'}\phi\phi'\delta(d_g(\xi,\xi')-R) 
\eea
whereas the moments $\vev{L^n}(R,V)$ are defined in terms of the loop
length distribution function $\rho_V(R,L)$ which counts the number of
loops of length $L$ which compose the boundary of a geodesic sphere
of radius $R$
\beq{*4}
 \vev{L^n}(R,V)=\int\, dL\, L^n\, \rho_V(R,L)\, .
\eeq
Notice that $\vev{L^1}(R,V)\equiv n_1(R,V)$.  For pure gravity
$\rho_\infty(R,L)R^2$ is a function of only one scaling variable
$y=L/R^2$. This is a manifestation of the fractal structure of
space--time. It means that the structure of the boundary in two
gravity is the self similar at all distances and it shows that 
${\rm dim}[L]={\rm dim}[R^2]$. As a consequence one
finds that $\vev{L^n}(R,\infty)\sim R^{2n}$ for $n\ge 2$.  For $n=0,1$
$\vev{L^n}(R,V)$ picks up a cutoff dependence from the short distance
behaviour of $\rho_\infty(R,L)R^2$ giving $\vev{L^n}(R,\infty)\sim
R^{d_h-1}$, which defines the fractal dimension $d_h$. $d_h=4$ for
pure gravity. For finite volume $V$ one expects a diverging
correlation length as one tunes the cosmological constant to its
critical value. Its size for pure gravity turns out to be $\xi_G\sim
V^{1/d_H}$ where $d_H$ is another definition of the fractal
dimension. It is a non trivial fact that $d_H=d_h$ for pure
gravity. In the case where $c\le 1$ matter is coupled to gravity, numerical
simulations support that the properties of the space--time geometry
are quite similar, except that the fractal dimension could be
different. Then from scaling arguments one expects that
\bea{*5}
\vev{L^{0,1}}(R,V)&=& V^{1-1/d_H} F_{0,1}(x)\, ,\\
\vev{L^n}    (R,V)&=& V^{2n/d_H}  F_n    (x)\, ,\quad n>1\, ,\\
n_\phi       (R,V)&=& V^{1-1/d_H-\Delta} F_\phi(x)\, \\
G_\phi       (R,V)&=& V^{-\Delta} g_\phi(x)\, .  
\eea
$x$ is the scaling variable $R/V^{1/d_H}$ and $G_\phi(R,V)$ is the
normalized matter two point function. For $x\ll 1$ one expects
$F_{0,1}(x)\sim x^{d_h-1}$, $F_n(x)\sim x^{2n}$, $F_\phi(x)\sim
x^{d_h-1-\Delta d_h}$ and $g_\phi(x)\sim x^{-\Delta d_h}$.
In our simulations the volume
$V$ is the number of triangles $N$, we use the
link distance $r$ \cite{syracuse} and we add the so called shift $a$ as a
finite size correction to $x$, {\it i.e.} $x=(r+a)/N^{1/d_H}$.

\centerline{\epsfxsize=7.0cm \epsfysize=4.67cm \epsfbox{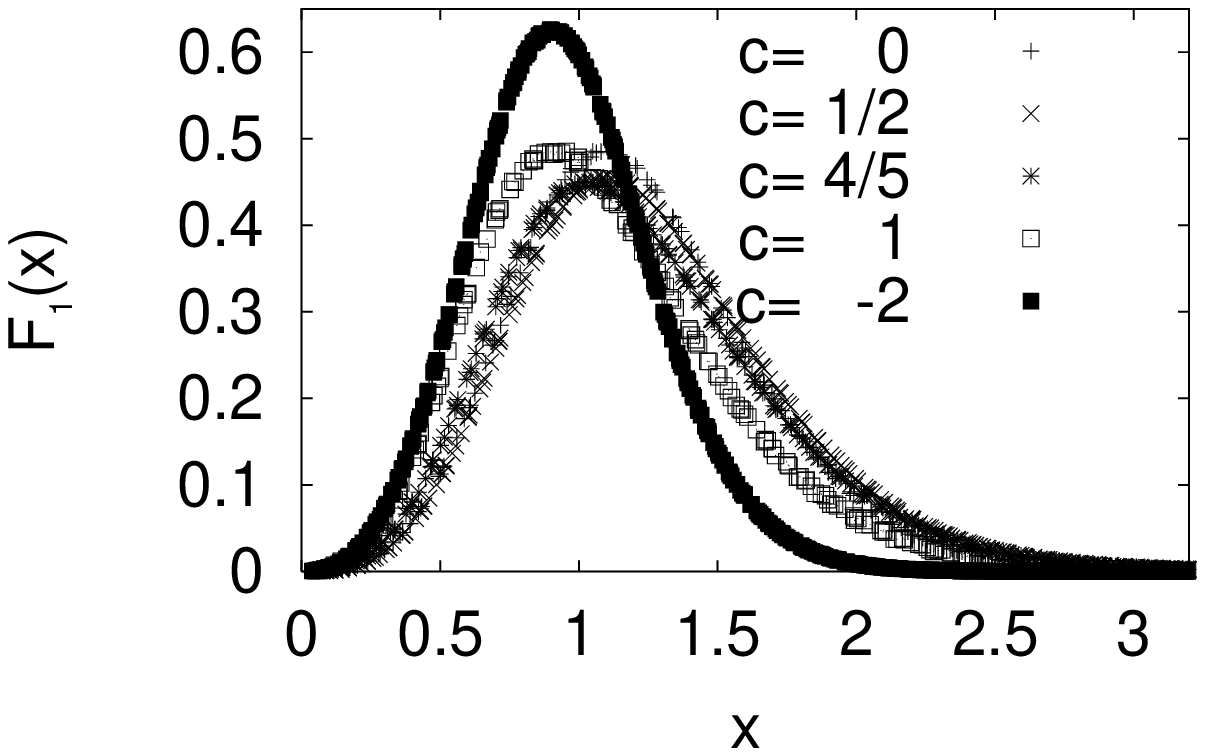}}

\vspace{-2mm}
\noindent {\bf Figure 1.} Collapse of the two point function
$\vev{L^1}(R,V)$ according to eq.~\protect\rf{*5}.

\centerline{\epsfxsize=7.0cm \epsfysize=4.67cm \epsfbox{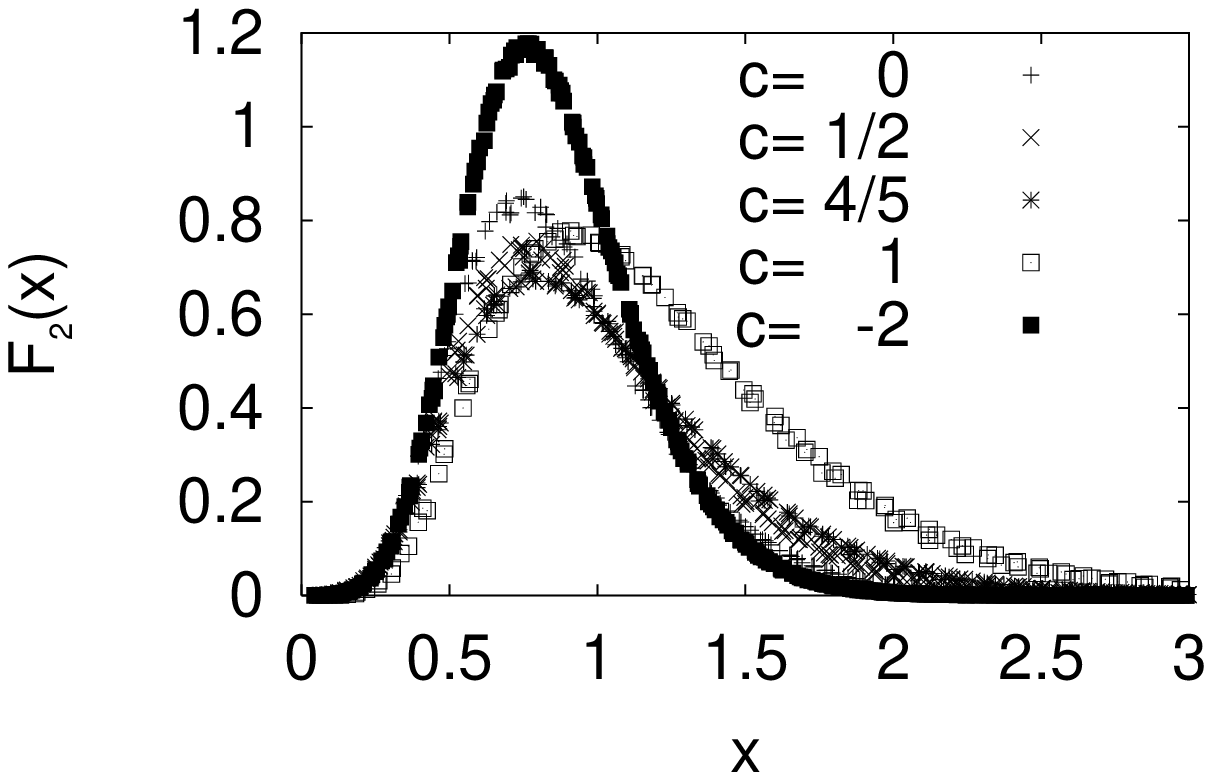}}

\vspace{-2mm}
\noindent {\bf Figure 2.} Collapse of 
$\vev{L^2}(R,V)$ according to eq.~\protect\rf{*5}.

\vspace{1mm}

\noindent
$d_H$
and $a$ are chosen so as to minimize the $\chi^2$ associated with the
distance between the ciruves for different $N$ according to
\cite{syracuse}. 

Lattice sizes up to 8,192K ($c=-2$), 128K ($c=0$), 256K
($c=1/2, 4/5$) and 64K ($c=1$) were used in the calculations.
The results are shown in Table~\ref{t:1} and in Figs. 1 and 2.
In Table~\ref{t:1}, FSS refers to finite size scaling according to
Eq.~\rf{*5} and SDS to small distance scaling, {\it i.e.} fits to the
expected small $x$ behaviour of the scaling functions $F_n(x)$ and
$g_\phi(x)$.  

The $c\le 1$ fractal properties of space--time disappear when we go
deep in the branched polymer phase ($c\ge 5$). We still have a
diverging correlation length $\xi_G\sim V^{1/d_H}$ when we tune the
cosmological constant to its critical value where now $d_H\approx
2$. We also find that $d_H\approx d_h$. 
The scaling variable $x$ still exists over all distance scales and
we can use finite size scaling in order to determine $d_H$. The loops
on the boundary never grow larger than the lattice cuttoff
$\varepsilon$. We observe that $\vev{L^n}(R,V)\sim V^{1/2} F_n(x)$
{\it independently of} $n$. For $c=8$ the actual {\it values} of
$\vev{L^n}(R,V)$ are independent of $n$ for given $V$. We also observe
that the maximum loop size is almost constant with $V$ whereas it
grows as $V^{2/d_H}$ for $c\le 1$. Therefore the loop length
distribution function has the expected branched polymer bahaviour
$\rho_\infty(R,L)= R\,\delta(L-\varepsilon)$.

\centerline{\epsfxsize=7.0cm \epsfysize=4.67cm \epsfbox{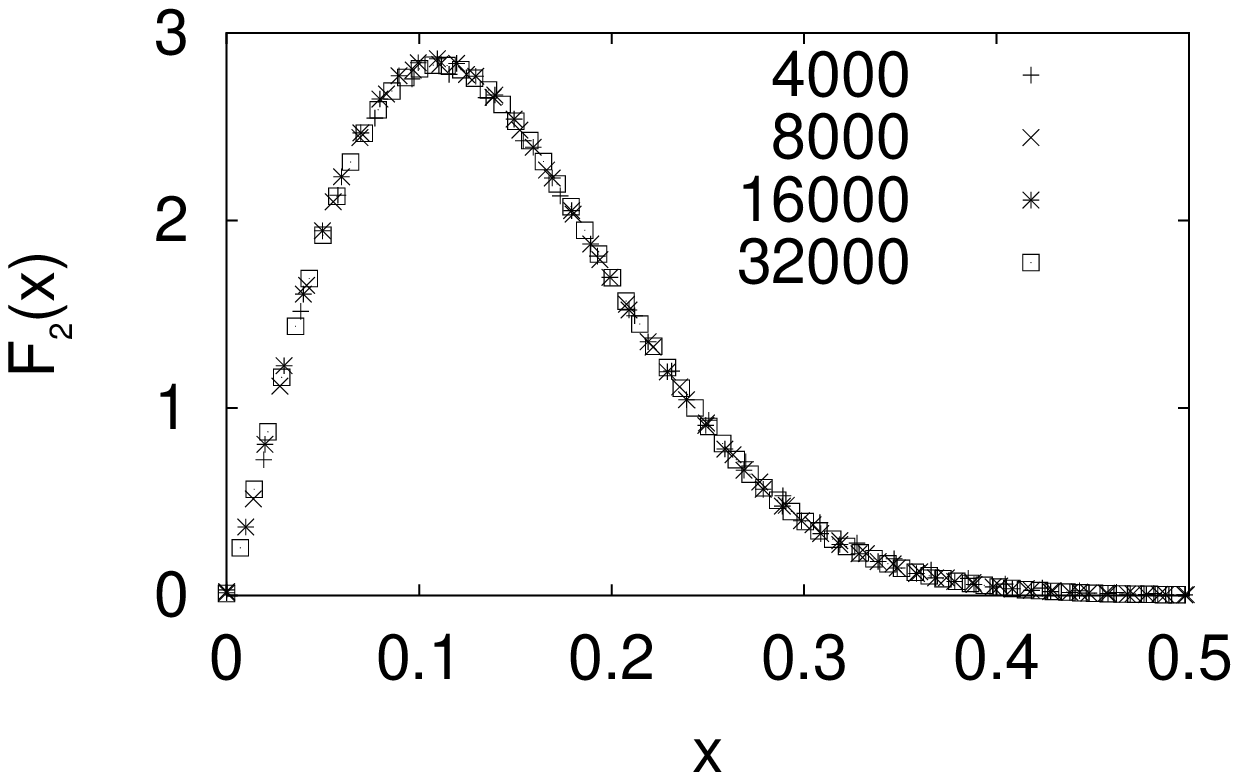}}

\vspace{-2mm}
\noindent {\bf Figure 3.} Collapse of 
$\vev{L^2}(R,V)\sim V^{1/2} F_2(x)$ for $c=5$.

\end{document}